\documentclass{cimento}

\usepackage{graphicx}  % got figures? uncomment this
\usepackage[applemac]{inputenc}
\usepackage[T1]{fontenc}
\usepackage[english]{babel}
\usepackage{amsmath}

\title{The rotation angle distribution underlying magnetic field fluctuations in the 1/{\it f} range of solar wind turbulent spectra}
\author{L.~Matteini\from{ins:x}\from{ins:y}\ETC,
D.~Stansby\from{ins:y},
T.S.~Horbury\from{ins:y}
        \atque
C.H.K.~Chen\from{ins:z}\from{ins:y}}%\thanks{Any footnote to author.}
\instlist{\inst{ins:x} LESIA, Observatoire de Paris, Université PSL, CNRS, Sorbonne Université, Univ. Paris Diderot, Sorbonne Paris Cité, 5 place Jules Janssen, 92195 Meudon, France
  \inst{ins:y} Department of Physics, Imperial College London, SW7 2AZ London, UK
  \inst{ins:z}{School of Physics and Astronomy, Queen Mary University of London, London E1 4NS, UK}
}
  
\shorttitle{Magnetic rotations in the 1/{\it f} solar wind spectrum}

\begin{document}

\maketitle

\begin{abstract}
We discuss properties of large amplitude magnetic field fluctuations during fast Alfvénic solar wind streams, focussing on the statistics of the rotation angle between consecutive magnetic field vector measurements for different scales in the plasma. Since in the fast solar wind fluctuations preserve the modulus of the magnetic field to a good approximation, the tip of the magnetic field vector is observed to move on a sphere of approximately constant radius $|\bf{B}|$. We then compare statistics of solar wind measurements with that of a simple model of a random walk bounded on a spherical surface. The analogy consists in the fact that in both systems the geometrical constraint imposes a limiting amplitude at large separations and thus introduces a break scale in the power spectrum of the fluctuations, leading to a shallower slope for scales where the fluctuations amplitude becomes scale-independent. 
However, while in the case of the random walk the saturation of the fluctuations occurs when the pattern becomes uniform on the sphere (flat distribution of the cosine of the rotation angle), transitioning then to a white noise regime, in the solar wind magnetic field fluctuations saturate in amplitude maintaining a preferential direction. We suggest that this behaviour is due to the presence of the background interplanetary magnetic field, which keeps some long-range memory in the system also when the fluctuations becomes independent of the scale. This long-range correlation is a necessary ingredient in order to produce the 1/{\it f} spectrum observed at large scales in the solar wind.
\end{abstract}

\section{Introduction}
In a recent paper \cite{Matteini_al_2018} we have discussed properties of the large scale solar wind magnetic spectrum, focussing on the so called 1/{\it f} range. This is the portion of the power spectrum of the fluctuations which corresponds to scales above the turbulent Kolmogorov inertial range and during fast streams it is characterised by a $-1$ spectral slope \cite{Bavassano_al_1982, Denskat_Neubauer_1982}. The origin of this range and its spectral shape are still debated \cite{Matthaeus_Goldstein_1986, Velli_al_1989, Verdini_al_2012, Chandran_2018}.
We briefly summarise here the main idea behind our model: in magnetised plasmas, if imposing a regime of low magnetic compressibility (as typically observed in the fast solar wind) we expect that there is a limiting amplitude for the fluctuations $\delta \bf{B}$ when they reach order of the background field $\bf{B}$; indeed, if $|\delta \bf{B}|\gg |B|$ fluctuations in the components induce also changes in the magnetic field intensity, so that they become highly compressible.
Note that in the fast solar wind, at large scales $|\delta {\bf B}|/|{\bf B}|\sim1$ but $\delta |\bf{B}|\ll |\delta \bf{B}|$, so that the fluctuations are nearly incompressible ($|\bf{B}|$ is conserved).
In \cite{Matteini_al_2018} we suggest that in a turbulent plasma, when the amplitude of the fluctuations becomes large so that at some scale $l_0$ $|\delta {\bf B}|\sim |\bf{B}|$, a saturation occurs to prevent the their further growth at scales $l>l_0$, and thus maintaining the plasma at a low level of magnetic compressibility. This leads to a break in the spectrum at $l_0$ and a constant level $|\delta \bf{B}|/|\bf{B}|\sim1$ for scales $l>l_0$.
Remarkably, this simple phenomenological scenario applies well to the large amplitude fluctuations of the solar wind; moreover in situ spacecraft measurements show that the scale $l_0$ at which $|\delta \bf{B}|/|\bf{B}|$ reaches order unity in the solar wind plasma always identifies the spectral break between 1/{\it f} and inertial ranges as observed at various radial distances from the Sun \cite{Matteini_al_2018}.

 An important comment is in order here. The existence of a flat first order structure function over some scales (constant $|\delta \bf{B}|$) does not directly imply a unique spectral slope. For example, if the fluctuations of this range are completely uncorrelated (white noise), the associated power spectrum is also flat.
On the other hand, when the autocorrelation function of the fluctuations is not identically zero (some long range correlation exists in the system), then the usual connection between second order structure functions at scale $l$, $\delta B_l^2$, and power spectral density (PSD) for k-vectors $k=1/l$, $P(k)$, can be made:
\begin{equation}\label{eq_1}
\delta B_l^2=P(k)\cdot k
\end{equation}
so that if the former has slope $\alpha$, the latter has slope $\alpha-1$ \cite{Monin_Yaglom_1975}.
It is then straightforward to see that our model, based on the saturation of the structure function $\delta B_l$ above a scale $l_0$ ($\alpha=0$), directly predicts a $-1$ spectrum. Also note that for eq.~\ref{eq_1}, $-1$ is the steepest possible slope.

However, as mentioned, this is possible only if some long-range correlations are present in the plasma at large scales \cite{Keshner_1982}, otherwise a flat (white noise) spectrum would be expected. The aim of this paper is then to investigate further this aspect in solar wind observations, and compare the result with a simple model of a bounded random walk.

\section{A bounded random walk model}
The fact that a self-similar process, which is in principle scale invariant, may be characterised by  a transition between different regimes at a particular scale, should not be surprising.
Consider in fact random walks on a 2-D surface; the two-point correlation for a random walk ensemble leads to a power law distribution with slope $-2$. This is however true if the walk takes place on an unbounded surface, so that separations  at arbitrary large scales can be performed. The large scale correlation in this case preserves the same properties as the small scale ones, leading to a single power-law spectrum. 
On the other hand, consider now the case when the random walk develops on a spherical surface. While on distances (separation times) that are small compared to the curvature radius of the surface 2-point correlations still display the same $-2$ spectrum as in the unbounded case, large scale correlations are characterised by a different behaviour. Indeed, if the random walk is let evolve for sufficiently long time, the pattern starts to cover uniformly the sphere, and the trajectory will eventually come back to close to the starting point. 

\begin{figure}
\includegraphics[width=13.5cm]{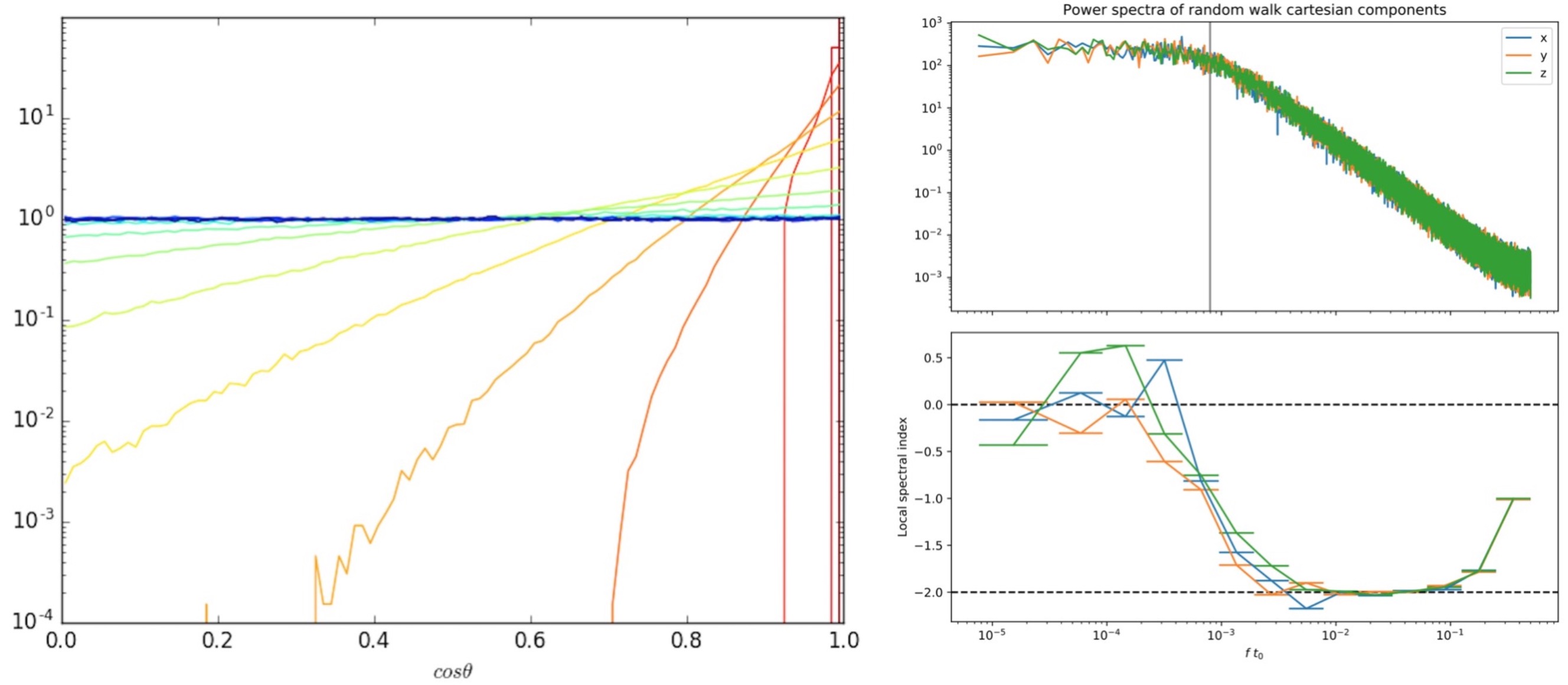}    % includes figure foo.eps
\caption{Results from a random walk model bounded on a spherical surface. Left: PDF of the cosine of the rotation angle $\theta$ between 2 vectors at different time separations encoded by colours, lag increasing from red to blue. 
Right: Power spectrum of the resulting time series of the random walk displacements on the sphere (top) and associated local spectral slopes (bottom). The expected break scale, corresponding to the number of steps needed to cover uniformly the sphere, is indicated by the solid line the in top panel.}
\label{fig_walk}
\end{figure}

An example of this dynamics is shown in fig.~\ref{fig_walk} where we report results from a simple model \cite{Barnes_1981} which tracks the motion of a vector whose tip is constrained on a sphere and subject to a random walk. The model is let evolve for $2^{17}$ time steps of $\Delta\theta=0.1$rad, and results are taken as averages of 10 ensembles.
The left panel shows the PDFs of the rotation angle $\theta$ between two distinct vectors for increasing time lags (i.e. number of steps $N$), from red to dark blue; at sufficiently large time lags, the distribution of $\cos{(\theta)}$ becomes flat, meaning a uniform coverage of the sphere, as expected, and does not evolve further when increasing the time lag.
This suggests that, although the underlying process generating the pattern is self-similar, the statistics made on time separations that are larger than the typical time needed to walk around the full sphere is different from that of smaller scale. Large scales are influenced by the fact that the walk occurs on a closed surface, and once this uniformly covered, they follow a different law.

This behaviour obviously affects the spectral properties of the system.
In the right panel of fig.~\ref{fig_walk} we show the power spectrum of the Cartesian components of the resulting separation time series. We can see that at short time scales a spectral slope of index $-2$ is recovered, as expected for a pure random walk; on the contrary, at a scale of $N\sim10^{3}$ time steps $t_0$, corresponding approximately to the time needed to cover the whole sphere ($N\sim4\pi/\Delta\theta^2$) and indicated by the vertical solid line, there is a change in the slope leading to a flat spectrum, as expected for vanishing correlation between points uniformly distributed (i.e. white noise).

The one above is just a simple example of how the presence of a geometrical constraint can lead to a break in the self-similarity and thus introduce a change in the expected power law of the fluctuations. Although it does not apply directly to turbulent interplanetary magnetic field fluctuations, which cannot be modelled as a simple random walk, this is still an instructive example to compare with the solar wind observations described in the next section.

\section{Solar wind observations}
Let us now consider in situ observations from the solar wind. Following \cite{Matteini_al_2018} we use Ulysses measurements at high latitudes, when the spacecraft was continuously embedded in fast polar wind (i.e. very regular, highly Alfvénic and with little variations of the magnetic and plasma pressures).

\begin{figure}
\includegraphics[width=13.5cm]{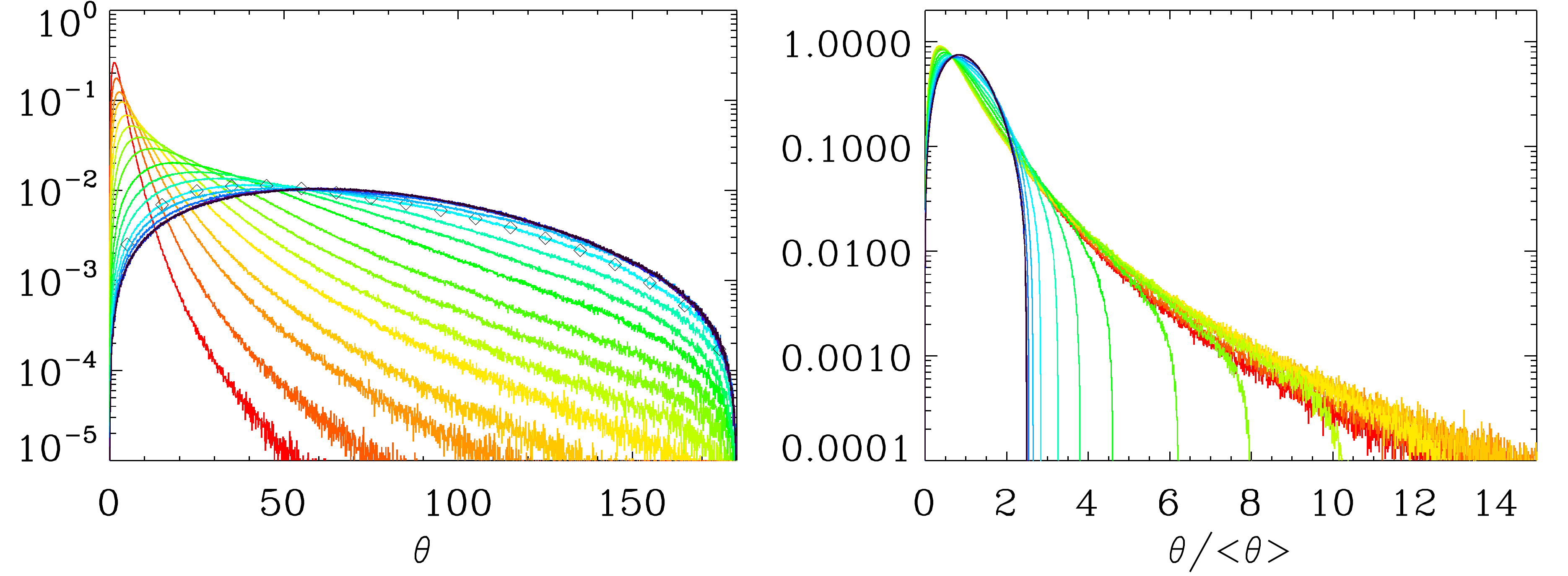} 
\caption{PDF of the rotation angle $\theta$ between 2 magnetic field vector measurements separated by a variable time lag $\Delta t$ in the fast solar wind. Colours encode different scales, from small (red) to large (blue) separations. Left and right panels show the distribution of $\theta$ and normalised $\theta/\left<\theta\right>$, respectively. As a reference, red/green PDFs belong to the turbulence inertial range (${1<\Delta t(\rm{s})<10^3}$) while almost overlapping blue distributions correspond to the 1/{\it f} range (${\Delta t>5\cdot10^3}$s). The PDF approximatively corresponding to the break scale separating the two ranges is highlighted with black diamonds}
\label{fig_alpha}
\end{figure}

We focus on the change of orientation between two different magnetic field measurements $B(t)$ and $B(t+\Delta t)$ as a function of the scale $\Delta t$. We recall that due to the high speed of the flow, in the solar wind time measurements (frequencies) really correspond to Doppler-shifted spatial scales (k-vectors).
Following \cite{Chen_al_2015} we calculate the PDFs of the angle $\theta$ between the two magnetic field vectors for different time lags $\Delta t$ (see also \cite{Matteini_al_2018} for details).
The left panel of fig.~\ref{fig_alpha} shows the distribution of magnetic rotations in the fast solar wind corresponding to Ulysses measurements at approximately 2 AU, from day 100 to 250 of year 1995. Colours encode different scales, from small (red/orange) to large (blue/purple) scales; the PDF approximatively corresponding to the break between inertial and 1/{\it f} ranges ($\Delta t\sim 5\cdot10^3$s) is highlighted with black diamonds. Due to the small power in the fluctuations at small scales the PDF of $\theta$ for small time lags peaks at small angles. Rotations become significantly larger in the inertial range of the turbulence (green curves), starting to reach the extreme boundary of 180 degrees. At larger scales, corresponding to the 1/{\it f} range (blue/purple), the PDF approaches a quasi-symmetric shape between 0 and 180 degrees; moreover, PDFs do not evolve further, as expected for scale-independent fluctuations (consistent with that, the corresponding PDFs of the fluctuations $\delta B$ also do not change inside the 1/{\it f} range and saturate to a constant level $|\delta {\bf B}|/|{\bf B}|\sim1$, see \cite{Matteini_al_2018}). 

The right panel of fig.~\ref{fig_alpha} shows the same distributions each normalised to its mean value $\left<\theta\right>$. As discussed in \cite{Chen_al_2015}, at small scales the normalised PDFs can be reasonably well described by a lognormal distribution; at these scales the PDFs of $\theta$ (left panel) are sufficiently far from the 180 degrees boundary, so that a high-tail can be sustained. However, moving to larger scales and approaching the 1/{\it f} range, the effect of the boundary on the lognormal tail is apparent. As a consequence, the distribution of $\theta/\left<\theta\right>$ in the 1/{\it f} has a more Gaussian-like shape, although some skewness is visible.

\begin{figure}
\includegraphics[width=13.5cm]{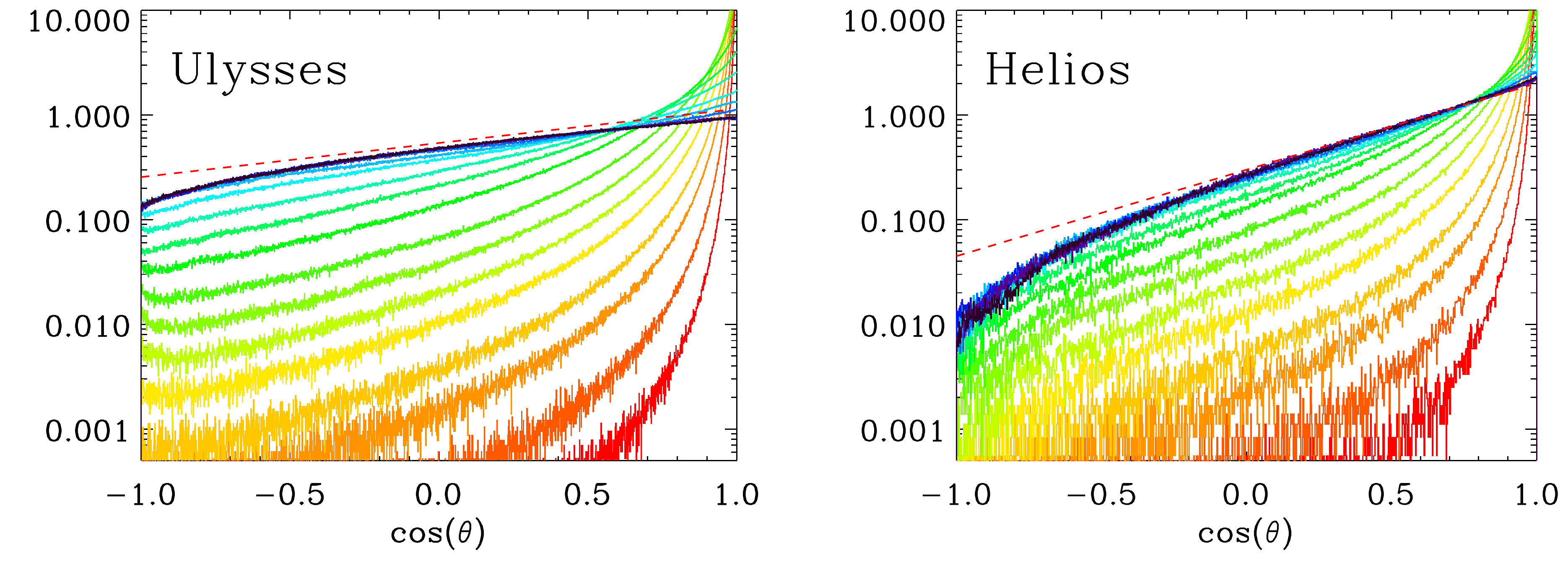}    % includes figure foo.eps
\caption{Left: PDF of the cosine of the rotation angle $\theta$ between 2 magnetic field vectors in the fast solar wind at 2 AU (Ulysses); colors encode different scales as in previous figure. At the largest scales (1/{\it f} range, blue) the distribution tends to become shallower, with a roughly constant slope, although not entirely flat; this implies that the distribution of the underlying magnetic field vector tends to spread over the full sphere of radius $|\bf{B}|$, but not yet uniformly. Right: same analysis for the fast wind at 0.3 AU (Helios measurements).}
\label{fig_cos}
\end{figure}

In order to better appreciate the spread of rotation angles, the left panel of fig.~\ref{fig_cos} shows the distribution of $\cos{\theta}$ for all scales, with the same colour code as fig.~\ref{fig_alpha}. This goes from a strongly peaked shape at small scales -where angles are almost always smaller than a few degrees- to shallower distributions at larger scales. Once more, note that curves corresponding to the 1/{\it f} range all show the same shape and fall on top of each other.
However, unlike fig.~\ref{fig_walk}, the PDFs never reach a flat shape, indicative of a uniform distribution over the sphere of constant $|B|$; indeed, a clear asymmetry between positive and negative values of $\cos{\theta}$ is maintained at all scales, including the 1/{\it f} range. As a consequence, rotations are preferentially distributed around the direction of the background field (approximately the Parker spiral). This implies that even at the largest scales (in the 1/{\it f} range, where $|\delta {\bf B}|/|{\bf B}|\sim1$) the fluctuations are not totally uncorrelated and have some long-range correlation, the memory of the background field direction.

In \cite{Matteini_al_2018} we have shown that the saturated shape observed for the 1/{\it f} range, corresponds to an approximatively exponential relation:
\begin{equation}\label{eq1}
PDF(\cos{\theta})\sim \exp{(\alpha\cos{\theta})}
\end{equation}
where $\alpha$ is an empirical constant, whose value can be obtained from the observations. 
Relation (\ref{eq1}) is shown in fig.~\ref{fig_cos} as a dashed red line; it seems to describe reasonably well the main part of the PDF in the 1/{\it f} range. For the Ulysses observations, we get $\alpha\sim0.8$.
As a comparison, we show in the right panel of the same figure the same analysis for a fast solar wind stream observed by the Helios spacecraft at 0.3 AU (see \cite{Matteini_al_2018} for more details). Also in this case we recover the same qualitative picture as for Ulysses (taken at $\sim2$AU); for Helios however, the PDFs in the 1/{\it f} are less shallow, and $\alpha\sim1.8$. This suggests a possible evolution of the PDF saturated shape with increasing radial distance.

\section{Discussion and conclusion}
To summarise, we have discussed properties of large scales fluctuations in the solar wind, when $|\delta {\bf B}|/|{\bf B}|\sim1$ and the tip of magnetic field vector approximatively moves on a sphere of constant radius $|{\bf B}|$ \cite{Bruno_al_2004, Matteini_al_2015a}, and compared them to a simple model based on a random walk bounded on a sphere. In particular we have focussed on the rotation angle $\theta$ between two different vectors separated by a variable time lag, to probe the effect  of the geometrical boundary on the statistics of different scales.

We have found that the difference in the two cases lies in the condition that characterises and constrains the asymptotic state.
In the simple random walk model, the self-similarity is broken by a purely geometrical condition, i.e. the pattern of the walk becomes uniform over the sphere. The appearance of a spectral break is then caused by the lack of two-point correlations above a certain scale (the size of the sphere) and as a consequence the associated spectrum becomes flat, i.e. white noise.

On the other hand, in the solar wind case some physical mechanism must be responsible for the saturation of the amplitude of the fluctuations typically observed at large scales \cite{Mariani_al_1978}; in \cite{Matteini_al_2018} we propose that this is related to a constraint of low magnetic compressibility in the plasma. As for the bounded random walk, the saturation of the amplitude of the fluctuations naturally introduces a break in the solar wind magnetic spectrum.
However, unlike fig.~\ref{fig_walk}, the distribution of $\cos{\theta}$ in the fast solar wind never reaches a flat shape (fig.~\ref{fig_cos}), suggesting that a preferential direction common to all fluctuations continue to exist: this is the interplanetary background magnetic field. This maintains some level of long-range correlation between the fluctuations, as required to build a 1/{\it f} spectrum like that observed in the solar wind at large scales.

\acknowledgments
%This work was supported by the Programme National PNST of CNRS/INSU co-funded by CNES.
This work was supported by national program PNST of CNRS/INSU co-funded by CNES. 
DS and TH are supported by STFC grants ST/K001051/1 and ST/N000692/1, respectively. CHKC is supported by STFC Ernest Rutherford Fellowship ST/N003748/2.

%\bibliographystyle{varenna}
%\bibliography{aamnem99,mybib}

\end{document}